\documentclass[]{pasj01}

\begin{document} 
\Received{2016/12/02}
\Accepted{2016/12/29}

\title{The K2-ESPRINT Project VI: K2-105~b, \\a Hot-Neptune around a Metal-rich G-dwarf}

\author{Norio \textsc{Narita}\altaffilmark{1,2,3}}%
\author{Teruyuki \textsc{Hirano},\altaffilmark{4}}
\author{Akihiko \textsc{Fukui}\altaffilmark{5}}
\author{Yasunori \textsc{Hori}\altaffilmark{2,3}}
\author{Fei \textsc{Dai}\altaffilmark{6}}
\author{Liang \textsc{Yu}\altaffilmark{6}}
\author{John \textsc{Livingston}\altaffilmark{1}}
\author{Tsuguru \textsc{Ryu}\altaffilmark{3,7}}
\author{Grzegorz \textsc{Nowak}\altaffilmark{8,9}}
\author{Masayuki \textsc{Kuzuhara}\altaffilmark{2,3,4}}
\author{Yoichi \textsc{Takeda}\altaffilmark{3}}
\author{Simon \textsc{Albrecht}\altaffilmark{10}}
\author{Tomoyuki \textsc{Kudo}\altaffilmark{11}}
\author{Nobuhiko \textsc{Kusakabe}\altaffilmark{2,3}}
\author{Enric \textsc{Palle}\altaffilmark{8}}
\author{Ignasi \textsc{Ribas}\altaffilmark{12}}
\author{Motohide \textsc{Tamura}\altaffilmark{1,2,3}}
\author{Vincent \textsc{Van Eylen}\altaffilmark{13}}
\author{Joshua N. \textsc{Winn}\altaffilmark{14}}

\altaffiltext{1}{Department of Astronomy, The University of Tokyo, 7-3-1 Hongo, Bunkyo-ku, Tokyo 113-0033, Japan}
\altaffiltext{2}{Astrobiology Center, NINS, 2-21-1 Osawa, Mitaka, Tokyo 181-8588, Japan}
\altaffiltext{3}{National Astronomical Observatory of Japan, NINS, 2-21-1 Osawa, Mitaka, Tokyo 181-8588, Japan}
\altaffiltext{4}{Department of Earth and Planetary Sciences, Tokyo Institute of Technology,
2-12-1 Ookayama, Meguro-ku, Tokyo 152- 8551, Japan}
\altaffiltext{5}{Okayama Astrophysical Observatory, National Astronomical Observatory of Japan,
NINS, Asakuchi, Okayama 719-0232, Japan}
\altaffiltext{6}{Department of Physics, and Kavli Institute for Astrophysics and Space Research,
Massachusetts Institute of Technology, Cambridge, MA 02139, USA}
\altaffiltext{7}{SOKENDAI (The Graduate University for Advanced Studies), 2-21-1 Osawa, Mitaka, Tokyo 181-8588, Japan}
\altaffiltext{8}{Instituto de Astrof\'{i}sica de Canarias (IAC), 38205 La Laguna, Tenerife, Spain}
\altaffiltext{9}{Departamento de Astrof\'{i}sica, Universidad de La Laguna (ULL), 38206 La Laguna, Tenerife, Spain}
\altaffiltext{10}{Stellar Astrophysics Centre, Department of Physics and Astronomy,
Aarhus University, Ny Munkegade 120, DK-8000 Aarhus C, Denmark}
\altaffiltext{11}{Subaru Telescope, National Astronomical Observatory of Japan, 650 North Aohoku Place, Hilo, HI 96720, USA}
\altaffiltext{12}{Institut de Ci\`{e}ncies de l'Espai (CSIC-IEEC), Carrer de Can Magrans, Campus UAB, 08193 Bellaterra, Spain}
\altaffiltext{13}{Leiden Observatory, Leiden University, 2333CA Leiden, Netherlands}
\altaffiltext{14}{Department of Astrophysical Sciences, Princeton University, Princeton, NJ 08544}

\email{narita@astron.s.u-tokyo.ac.jp}



\maketitle

\begin{abstract}
We report on the confirmation that the candidate transits observed for the star
EPIC~211525389 are due to a short-period Neptune-sized planet.
The host star, located in {\it K2} campaign field~5,
is a metal-rich ($[\mathrm{Fe/H}] = 0.26\pm0.05$)
G-dwarf ($T_{\rm eff} = 5430 \pm 70$ K and $\log g = 4.48 \pm 0.09$),
based on observations with the High Dispersion Spectrograph (HDS)
on the Subaru 8.2m telescope.
High-spatial resolution AO imaging with HiCIAO on the Subaru telescope
excludes faint companions near the host star, and the false positive 
probability of this target is found to be $<$$10^{-6}$ using the open source \texttt{vespa} code.
A joint analysis of transit light curves from {\it K2} and additional ground-based
multi-color transit photometry with MuSCAT on the Okayama 1.88m telescope
gives the orbital period of $P  = 8.266902 \pm 0.000070$ days and 
consistent transit depths of $R_p/R_\star \sim 0.035$ or $(R_p/R_\star)^2 \sim 0.0012$.
The transit depth corresponds to a planetary radius of
$R_p  = 3.59_{-0.39}^{+0.44}$ $R_{\oplus}$,
indicating that EPIC~211525389~b is a short-period Neptune-sized planet.
Radial velocities of the host star, obtained with the Subaru HDS,
lead to a 3$\sigma$ upper limit of $90 M_{\oplus}$ ($0.00027 M_{\odot}$)
on the mass of EPIC~211525389~b, confirming its planetary nature.
We expect this planet, newly named K2-105~b, to be the subject of future studies to characterize its mass, atmosphere,
spin-orbit (mis)alignment, as well as investigate the possibility of additional planets in the system. 
\end{abstract}

\section{Introduction}

Transiting planets are especially valuable targets in exoplanet studies due to the 
fact that both their radius and mass can be determined.
Thanks to previous ground-based and space-based transit surveys,
thousands of transiting exoplanets have been discovered.
The {\it K2} mission \citep{2014PASP..126..398H} is currently continuing the legacy of
{\it Kepler} in providing dozens of interesting transiting exoplanet candidates
in each of its successive $\sim 80$ day observing campaigns in the ecliptic plane.
Since 2014, {\it K2} has discovered  more than 100 new transiting exoplanets by
several follow-up teams (e.g., \cite{2015ApJ...804...10C,2015ApJ...812..112S,
2016arXiv160107635L,2016ApJ...818...46M,2016ApJS..226....7C}).

The large number of confirmed transiting exoplanets provide us
a unique opportunity to investigate their distribution in parameter space, such as 
the Period-Radius (P-R) relation, the Period-Mass (P-M) relation, and 
the Mass-Radius (M-R) relation.
While the M-R relation is useful to investigate the composition and
existence of volatile-rich atmosphere \citep{2013PASP..125..227Z},
P-R and P-M relations are suggested to provide possible insights into
planet formation and the migration of short-period planets
\citep{2013ApJ...763...12B,2013A&A...560A..51A,2016MNRAS.455L..96H,2016A&A...589A..75M}.

In this regard, short-period Neptune-sized planets are especially interesting,
since such planets occupy a region of parameter space which corresponds to the proposed dearth
of short-period super-Neptune/sub-Jovian planets.
Moreover, since transiting exoplanets allow us to investigate their atmospheres via transmission spectroscopy,
to measure spin-orbit (mis)alignments via the Rossiter-McLaughlin effect or doppler tomography,
and to probe possible presence of outer planets via transit timing variations,
they will incubate further follow-up science cases and provide
additional clues to uncover formation and migration mechanisms of
short-period planets.

In this paper, we report the confirmation of a new transiting hot Neptune
around a metal-rich G-dwarf EPIC~211525389 (TYC 807-1019-1) in {\it K2} campaign field 5.
The host star is relatively bright ($m_{K_p} = 11.69$ mag) with colors of $B-V = 0.79$ and $V-J = 1.38$,
and is located at the distance of $\sim230$ pc according to {\it GAIA} parallax (see \S3.1).
This target was identified as an interesting candidate planet host by the 
international collaboration ESPRINT,
{\it Equipo de Seguimiento de Planetas Rocosos Intepretando sus Transitos},
\citep{2015ApJ...812..112S,2016ApJ...820...56V,2016arXiv160509180V,
2016ApJ...820...41H,2016ApJ...825...53H,2016ApJ...823..115D}.
Although the star was also reported to be a candidate planet host by
\citet{2016MNRAS.461.3399P} and \citet{2016arXiv160702339B},
we have confirmed the planetary nature of this object for the first time
via high dispersion spectroscopy, high-contrast AO imaging,
additional ground-based transit photometry,
and radial velocity (RV) measurements.
The planet is newly named as K2-105~b.

The rest of this paper is organized as follows.
Observations of {\it K2}, high dispersion spectroscopy, AO imaging, and additional
transit photometry as well as our reduction methods are described in \S2.1--2.4.
We present stellar parameters of EPIC~211525389 (hereafter K2-105)
from high dispersion spectroscopy in \S3.1,
a contrast curve around the host star from AO imaging in \S3.2,
a joint transit analysis with light curves from {\it K2} and
ground-based multi-color transit photometry in \S3.3,
and RVs and a corresponding upper limit on the mass of EPIC~211525389~b (hereafter K2-105~b) in \S3.4.
We confirm the planetary nature of K2-105~b based on the mass upper limit
and a statistical analysis using the \texttt{vespa} code \citep{2012ApJ...761....6M, 2015ascl.soft03011M}
in \S4.1.
We report an improved transit ephemeris and a hint of a possible transit timing variation
for K2-105~b in \S4.2.
We discuss the importance of a discovery of a new short-period hot-Neptune
from a theoretical point of view in \S4.3.
Finally, we summarize our findings in \S5.

\section{Observations and Reductions}

\subsection{K2 Photometry with the ESPRINT pipeline}

\begin{figure}[tp]
			\includegraphics[width=8.5cm]{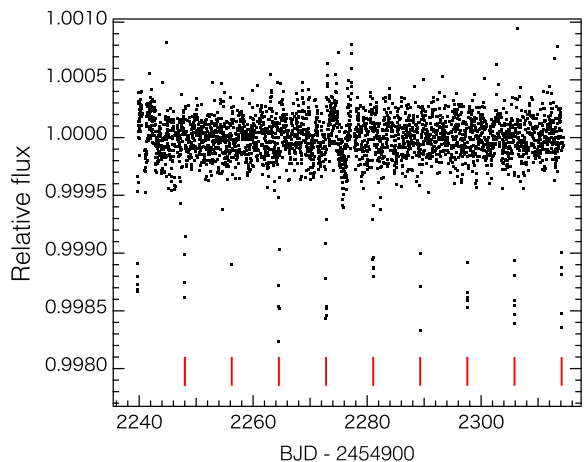} 
			\caption{K2 light curve of K2-105 processed by the ESPRINT pipeline.
			Red bars indicate positions of transits.
			\label{k2lightcurve}}
	\end{figure}

K2-105 was observed in {\it K2} campaign field 5
from 2015 April 27 to July 10.
We obtained the {\it K2} data from the
Mikulski Archive for Space Telescopes (MAST).
We found small variability in the raw light curve
with the amplitude of $\sim0.4$ \% and the period of $\sim24$ days,
possibly related with the stellar rotation.
We processed it with the ESPRINT pipeline \citep{2015ApJ...812..112S}
to create a detrended light curve (see Figure~\ref{k2lightcurve}).
In brief, we identified the candidate planet EPIC~211525389~b with
a Box-Least-Squares routine \citep{2002A&A...391..369K, 2010ApJ...713L..87J}
using the optimal frequency sampling described by \citet{2014A&A...561A.138O}.
The transit-like dimming occurred every $\sim$8.2672478 days with a depth of $\sim0.1$ \%.
No odd-even difference was observed in the dimming and
no significant evidence of secondary eclipses were seen, suggesting that
the signal is likely to be caused by a transiting planet.
We thus added this object as one of our follow-up targets.

\subsection{Subaru 8.2m Telescope / HDS}

We observed K2-105 with the High Dispersion Spectrograph
(HDS: \cite{2002PASJ...54..855N}) on the Subaru 8.2m telescope
located at the summit of Mauna Kea, Hawaii.
We employed the std-I2a setup
covering 4940\AA\ -- 7590\AA\
with the Image Slicer \#2
($R\sim80,000$: \cite{2012PASJ...64...77T}).
We took a template spectrum without an iodine cell on 2015 November 28 (UT).
The total exposure time was 1,500 s and the typical signal-to-noise ratio
around 6000\AA\ was $\sim60$.
The standard IRAF\footnote{The Image Reduction and Analysis
  Facility (IRAF) is distributed by the National Optical Astronomy
  Observatory, which is operated by the Association of Universities
  for Research in Astronomy (AURA) under a cooperative agreement with
  the National Science Foundation.}
procedures for HDS were applied, including overscan subtraction, non-linearity correction,
bias subtraction, flat fielding, scattered light subtraction, aperture extraction,
and wavelength calibration using Th-Ar lines, resulting in a calibrated one-dimensional spectrum. 
We also took HDS spectra with the same setup and with the iodine cell
to monitor RVs of K2-105 on 2015 Nov 26-28, 2016 Feb 2, and 2016 Oct 12-14 (UT).

\subsection{Subaru 8.2m Telescope / HICIAO \& AO188}

High contrast, high spatial resolution images of K2-105 were taken
in the $H$ band with HiCIAO \citep{2006SPIE.6269E..28T}
in combination with AO188 (188 element curvature sensor adaptive optics system:
\cite{2008SPIE.7015E..25H}) on the Subaru telescope on 2015 December 30 (UT).
The field of view of HiCIAO is about $20'' \times 20''$.
We used the target itself as a natural guide star for AO188,
producing the typical AO-corrected full-width-at-half-maximum (FWHM) of $\sim0\farcs06$.
All of our observations were carried out in siderial tracking mode.
We acquired 10 object frames with an individual exposure time of 5 s,
using a 9.74\%-transmittance neutral density (ND) filter
that avoids saturation of the target star.
Those unsaturated frames were used to calibrate the contrast limit around the target.
In addition, we performed the observations without the ND filter
and took 60 object frames with 15 s exposure,
enabling us to search for the faint sources around the target.
The total exposure time was 15 min (50 s) without (with) the ND filter.

The ACORNS pipeline (see \cite{2013ApJ...764..183B})
was used to reduce the HiCIAO data as follows.
First we remove a characteristic stripe bias pattern \citep{2010SPIE.7735E..30S},
and then bad pixel and flat-field correction are performed.
In order to correct HiCIAO's field distortion,
we compare HiCIAO data of the M5 globular cluster taken during the same run
with the archival M5 images taken by Advanced Camera for Surveys (ACS)
on {\it Hubble Space Telescope}, based on the same way as explained in \cite{2013ApJ...764..183B}.
Finally, the plate scale is corrected to be 9.5 mas pixel$^{-1}$.

\subsection{Okayama 188cm Telescope / MuSCAT}

We obtained simultaneous multi-band transit photometry of the target 
on 2016 February 10 (UT) using
MuSCAT \citep{2015JATIS...1d5001N} on the 188cm telescope
at the Okayama Astrophysical Observatory in Japan.
The sky condition during our observation was free of clouds and moonlight (2-day-old moon),
but slightly hazy due to yellow dust.
MuSCAT has the capability of simultaneous three-band imaging with
Sloan Gen 2 filters ($g'_2$, $r'_2$, and $z_{s,2}$) and three CCD cameras,
each having $6\farcm1$$\times$$6\farcm1$ FOV, with a pixel scale of about $0\farcs358$.
The exposure time was set to 60 s for $g'_2$ and $z_{s,2}$ bands,
and 20 s for $r'_2$ band.
We defocused the telescope such that the FWHM
of the stellar point spread function (PSF) was kept around 24 ($g'_2$),
29 ($r'_2$), and 32 ($z_{s,2}$) pixels, respectively.
The observations were conducted during JD 2457428.95 - 2457429.20.

The observed images are dark-subtracted, flat-fielded, and corrected for 
non-linearity, separately for each CCD.
Aperture photometry is performed for the target and two brighter comparison stars in the field of view
(TYC 807-1069-1 hereafter C1, and TYC 807-1165-1 hereafter C2) using a customized pipeline \citep{2011PASJ...63..287F}.
The aperture radius for each band is chosen as 24 ($g'_2$), 26 ($r'_2$), and 28 ($z_{s,2}$) pixels
respectively so that the apparent root-mean-square (RMS)
for a fractional light curve of C1/C2 is minimized.
We check for possible systematic variability of the target and comparison stars
by making the fractional light curves of each combination.
We find that the fractional light curves of C1/C2 in $r'_2$ and $z_{s,2}$ bands smoothly
change in a linear manner with the deviation from a linear approximation of $\sim$0.1\%.
On the other hand, the fractional light curve of C1/C2 in $g'_2$ band shows
a strange systematic variation with the amplitude of $\sim$0.4\%.
Although we suspect the systematic variation is caused by
strong 2nd-order extinction of Earth's atmosphere,
we decide not to use $g'_2$ band light curve in the subsequent analysis,
since we cannot correct the variation with the observed data.
The total flux of C1+C2 is used as a comparison flux to the target.
We also find that the peak count and total flux of the target suddenly dropped
after JD 2457429.17, even though the FWHM did not change,
suggesting the sky transparency changed significantly at that time.
We thus confine usable data to around JD 2457428.95 - 2457429.17
in the subsequent analysis to avoid systematic errors.

\section{Analyses and Results}

\begin{table}[tp]
\caption{Stellar Parameters of K2-105}\label{startable}
\begin{tabular}{lc}
\hline
Parameter & Value \\\hline
\multicolumn{2}{l}{\it (Stellar Parameters)\footnotemark[$a$]} \\

RA (J2000.0) & 08:21:40.871 \\
Dec (J2000.0) & +13:29:51.08 \\
$m_{K_p}$ [mag] & $11.687$ \\
$m_{g'}$ [mag] \footnotemark[$b$] & $12.244\pm 0.001$ \\
$m_{r'}$ [mag] \footnotemark[$b$] & $11.656\pm 0.001$ \\
$m_{i'}$ [mag] \footnotemark[$b$] & $11.484\pm 0.001$ \\
$m_{z'}$ [mag] \footnotemark[$b$] & $11.419\pm 0.015$ \\
$m_J$ [mag] & $10.541\pm0.02$ \\
$m_H$ [mag] & $10.173\pm0.03$ \\
$m_{K_\mathrm{s}}$ [mag] & $10.091\pm0.02$ \\
$B-V$ [mag] & $0.79\pm 0.05$ \\
$V-J$ [mag] & $1.38\pm 0.05$ \\ 
\hline
\multicolumn{2}{l}{\it (Spectroscopic Parameters)} \\
$T_{\rm eff}$ [K] & $5434\pm 35$ (stat.) $\pm 40$ (sys.) \\
$\log g$ [dex] & $4.477\pm0.085$\\
$[\mathrm{Fe/H}]$ [dex] & $0.26\pm0.05$ \\
$\xi$ [km s$^{-1}$] & $0.21\pm 0.44$\\
$v \sin i$ [km s$^{-1}$] & $1.76\pm 0.86$\\
\hline
\multicolumn{2}{l}{\it (Derived Parameters)} \\
$M_\star$ [$M_\odot$] & $1.01\pm0.07$ \\
$R_\star$ [$R_\odot$] & $0.95_{-0.10}^{+0.11}$\\
$\rho_\star$ [$\rho_\odot$] & $1.19_{-0.32}^{+0.44}$ \\
$\rho_\star$ [g cm$^{-3}$] & $1.68_{-0.45}^{+0.62}$ \\
Distance [pc] \footnotemark[$c$] & $220 \pm 30$\\
Distance [pc] \footnotemark[$d$] & $233_{-23}^{+29}$ \\
Age [Gyr] & $\geq 0.6$\\
\hline
\end{tabular}
\begin{tabnote}
\hangindent6pt\noindent
\hbox to6pt{\footnotemark[$a$]\hss}\unskip Based on the EPIC, SDSS, UCAC4, and 2MASS Catalogs.
\hbox to6pt{\footnotemark[$b$]\hss}\unskip Based on the SDSS PSF magnitude.
\hbox to6pt{\footnotemark[$c$]\hss}\unskip Based on the 2MASS apparent magnitude and
the estimated absolute magnitudes for the stellar parameters.
\hbox to6pt{\footnotemark[$d$]\hss}\unskip Based on the parallax reported by {\it GAIA} Data Release 1.
\end{tabnote}
\end{table}

\subsection{Spectroscopic Parameters}

We perform a line-by-line analysis for the HDS spectrum
following the method described in \citet{2002PASJ...54..451T} and \citet{2005PASJ...57...27T}. 
By measuring the equivalent widths of Fe I and Fe II lines between 5000 \AA\ and 7400 \AA,
we estimate the stellar effective temperature $T_{\rm eff}$, the surface gravity $\log g$,
the metallicity [Fe/H], and the microturbulent velocity $\xi$ from the excitation and ionization equilibria.
Based on the estimated atmospheric parameters, we estimate the mass, radius,
and density of the host star using the empirical relations derived by \citet{2010A&ARv..18...67T}
from detached binaries.
Note that the empirical relations have uncertainties of 6 \% and 3 \% for
the mass and radius, respectively, and these uncertainties are taken into account.
We also take into account the fact that the effective temperature derived from
the excitation/ionization may have a systematic error of about 40 K
(see details in \cite{2010MNRAS.405.1907B,2014ApJ...783....9H}).
The estimated mass and radius are in good agreement with those based on
the Yonsei-Yale stellar-evolutionary model \citep{2001ApJS..136..417Y},
which we use to set a lower limit on the age of the host star of 0.6 Gyr.
To derive the stellar rotational velocity $v \sin i$, we generate the stellar intrinsic 
spectrum using ATLAS9 model (a plane-parallel stellar atmosphere model in LTE; \cite{1993KurCD..13.....K})
assuming the above derived atmospheric parameters, and convolve the model spectrum with
the rotation plus macroturbulence kernel \citep{2005oasp.book.....G} and the instrumental profile of Subaru/HDS.
Taking account of the intrinsic uncertainty in the macroturbulent velocity \citep{2012ApJ...756...66H},
we estimate $v \sin i$ to be $1.76 \pm 0.86$ km s$^{-1}$. 

The distance of the host star is estimated as $220 \pm 30$ pc
by comparing the absolute magnitudes based on the Dartmouth isochrones
\citep{2008ApJS..178...89D} for the above stellar parameters
with the apparent magnitudes in $JHK_s$ bands from
the 2MASS point source catalog \citep{2006AJ....131.1163S}.
In addition, recently {\it GAIA} Data Release 1 \citep{2016arXiv160904153G,2016arXiv160904303L}
reported the parallax of K2-105 as $4.288 \pm 0.467$ mas,
corresponding to the distance of $233_{-23}^{+29}$ pc.
These two estimates are in excellent agreement,
implying the spectroscopically-derived stellar parameters are reasonable.

Derived stellar parameters and their errors are summarized in Table~\ref{startable}.
As a result, we find that the host star K2-105 is a metal-rich G-dwarf.
This result is, however, inconsistent with \citet{2016ApJS..224....2H},
who reported EPIC~211525389 to be a metal-poor giant based on reduced proper motion and colors.
Such discrepancies are not uncommon, given the difficulty of metallicity determination
based only on broadband photometry.
The spectroscopic determination is more reliable.

\subsection{Excluding Faint Contaminants}

	\begin{figure}[tp]
			\includegraphics[width=6.5cm]{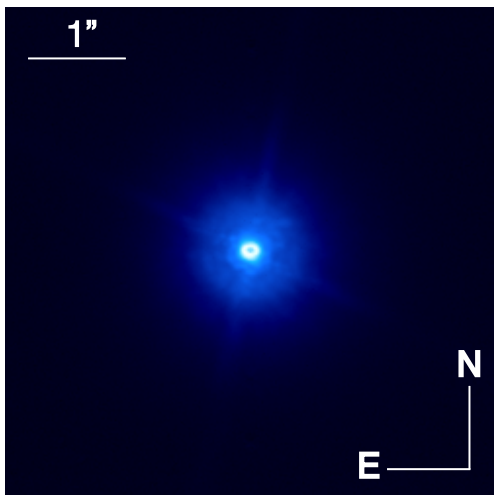} 
			\caption{A combined image around K2-105. The image is given in the log color scale.
			The central region around K2-105 is saturated. This figure shows a 5" $\times$ 5" portion.
			North is up and East is left.
			\label{aoimaging}}
	\end{figure}

	\begin{figure}[tp]
			\includegraphics[width=9cm]{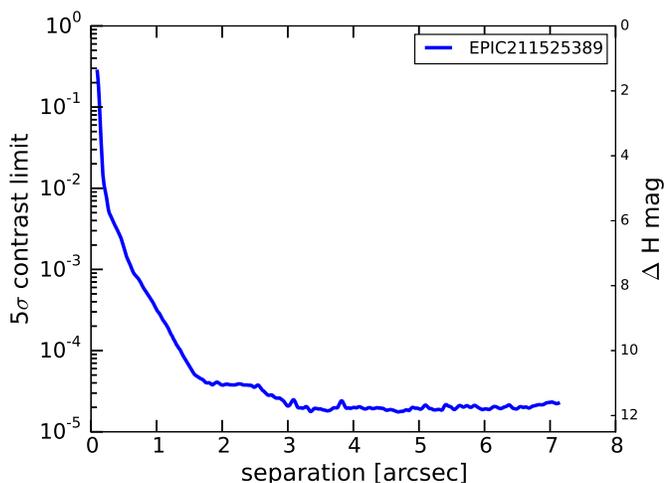} 
			\caption{A 5$\sigma$ contrast limit curve around K2-105.
			\label{contrastcurve}}
	\end{figure}

We compute offsets between the central star's centroids
in each frame obtained with and without the ND filter.
The reduced images of saturated and unsaturated frames are then offset-corrected,
sky-level-subtracted, and combined to produce final deep-integration images.
No additional point source which can mimic the observed transit signal is
detected in the final image.
Hence we compute the detection limit for such sources.
The combined saturated image is convolved with the FWHM of
the combined unsaturated image, and the standard deviations of
counts in annuli segmented from the center of the target are calculated.
Aperture photometry for the combined unsaturated image
is done to compute the flux count of the target per unit of time.
By comparing the flux of the target to the standard
deviations of counts in annuli, we create a 5$\sigma$ contrast curve
as a function of the separation in arcsec.
The combined image of the saturated frames and
the 5$\sigma$ contrast curve are shown in Figures~\ref{aoimaging} and \ref{contrastcurve}.
Consequently, we do not find any evidence of contaminants which
could mimic the transit signal around K2-105
at the level shown in the contrast curve.

\subsection{Joint Transit Light Curve Analysis}

	\begin{figure*}[tp]
			\includegraphics[width=17cm]{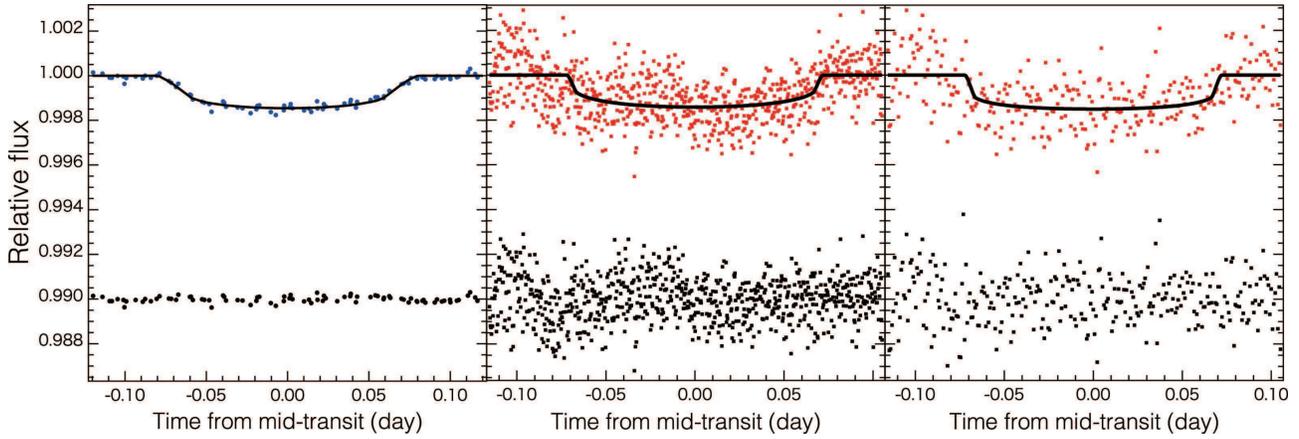} 
			\caption{Left panel: The phase-folded, baseline-corrected {\it K2} transit light curve.
			Blue dots plot the {\it K2} data. A black solid line represents
			the best-fit transit model that is integrated over the {\it K2} cadence
			($\sim 29.4$ minutes). Residuals from the best-fit model
			are plotted with a vertical offset of $-0.01$.
			Middle and right panels: Same as the left panel, but the baseline-corrected
			MuSCAT transit light curves for the $r'_2$ band (middle) and
			the $z_{s,2}$ band (right), respectively. Red dots show the MuSCAT data.
			\label{transitlc}}
	\end{figure*}

We simultaneously fit {\it K2} and MuSCAT transit light curves as follows.

The {\it K2} light curve shown in Figure~\ref{k2lightcurve} is separated into nine transit segments,
each containing a full-transit (namely, before, during, and after a transit).
Note that there is another transit at the beginning of {\it K2} observation,
but we exclude this transit since the data before the transit are not available.
Each segment includes the data within $\sim 9$ hours from
the apparent transit center.
We compute the standard deviation of the out-of-transit {\it K2} data,
excluding data in the transit segments and apparent outliers
(with the excursion larger than 0.001 from the unity).
We adopt the standard deviation of the out-of-transit data
as an estimate of the uncertainty for each {\it K2} flux value.
For {\it K2} transit light curves, we adopt a linear function in time as a baseline model,
\begin{eqnarray*}
F_{\rm base} (t) &=& k_0 + k_t t,
\end{eqnarray*}
where $k_0$ is the normalization factor and $k_t$ is the coefficient for the time.
Other parameters for the {\it K2} transit light curves are the mid-transit time
for each transit segment ($T_c (E)$, where $E=0$ -- 8),
and the planet-to-star radius ratio $R_p/R_\star$ for the {\it K2} bandpass ($K_p$).

For the MuSCAT transit light curves, we adopt a novel parametrization for the baseline model
to take account for the 2nd-order extinction introduced by \citet{2016arXiv161001333F}.
As a brief introduction, the parametrization uses the apparent magnitude of
the comparison star(s) instead of the airmass,
and the baseline function is expressed in magnitudes as follows,
\begin{eqnarray*}
m_{t, \mathrm{base}} (t) &=& k_0 + k_t t + k_c m_c (t),
\end{eqnarray*}
where $m_{t}$ and $m_c$ are the apparent magnitude of the target and comparison star(s),
$k_c$ is the coefficient for the atmospheric extinction.
This parametrization allows us to correct both the airmass extinction and
the 2nd-order extinction caused by the different spectral types of comparison stars.
See \citet{2016arXiv161001333F} for the mathematical derivation of this method.
Other parameters for the MuSCAT transit light curves are
the mid-transit time for the MuSCAT observation ($T_c (34)$) and
$R_p/R_\star$ for MuSCAT $r'_2$ and $z_{s, 2}$ bands.

Finally, the common planetary transit parameters are
the scaled semi-major axis $a/R_\star$ and
the impact parameter $b$.
We place an {\it a priori} constraint on $a/Rs$ to be $18.4 \pm 2.5$,
based on spectroscopically derived $M_\star$,  $R_\star$, and
the semi-major axis estimated by Kepler's third law.
We assume an orbital period of $P = 8.2672478$ days,
which we derive from the {\it K2} transits at the time of 
identification of the candidate.
Although we later derive an improved orbital period, the difference has
no impact on fitting results, since we allow all $T_c (E)$ to be free parameters.

To estimate the values of the free parameters and their uncertainties,
we use a code \citep{2007PASJ...59..763N}
that uses the analytic formula given by \citet{2009ApJ...690....1O} for the transit light curve model.
The analytic transit formula is equivalent to that given by \citet{2002ApJ...580L.171M}
when using the quadratic limb-darkening law.
We adopt a methodology for applying priors on limb-darkening described in \citet{2016ApJ...819...27F}.
To reduce a correlation between the quadratic limb-darkening coefficients
$u_1$ and $u_2$, and to appropriately estimate uncertainties for other parameters,
we use a combination form of the limb-darkening parametrization
$w_1 = u_1 \cos (\phi) - u_2 \sin (\phi)$ and
$w_2 = u_1 \sin (\phi) + u_2 \cos (\phi)$,
which was introduced by \citet{2008MNRAS.390..281P}.
We adopt $\phi = 40^{\circ}$ that is recommended by \citet{2008MNRAS.390..281P}
and \citet{2011MNRAS.418.1165H}.
We refer the tables of quadratic limb-darkening parameters by \citet{2013A&A...552A..16C}
and compute allowed $w_1$ and $w_2$ values for the stellar parameters presented
in Table~\ref{startable}.
We employ uniform priors for $w_1$ between [0.197, 0.359] for $K_p$ band,
[0.180, 0.329] for MuSCAT $r'_2$ band,
and [-0.028, 0.191] for MuSCAT $z_{s,2}$ band, respectively.
For $w_2$, we adopt Gaussian priors as $0.466\pm0.009$ for $K_p$ band,
$0.472\pm0.009$ for MuSCAT $r'_2$ band,
and $0.375\pm0.011$ for MuSCAT $z_{s,2}$ band, respectively.
The priors for the MCMC analysis are summarized in Table~\ref{priortable}.

\begin{table}[tb]
\caption{Summary of priors for the MCMC analysis.\label{priortable}}
\begin{tabular}{lll}
\hline
Parameter & Prior & Explanation  \\\hline
$P$ [days] & 8.2672478 & Fixed\\
$a/R_\star$ & $18.4\pm2.5$ & Added in the penalty function\\
$w_{1,K_p}$ & [0.197, 0.359] & Uniform prior\\
$w_{1,r'_2}$ & [0.180, 0.329] & Uniform prior\\
$w_{1,K_p}$ & [-0.028, 0.191] & Uniform prior\\
$w_{2,K_p}$ & $0.466\pm0.009$ & Gaussian prior\\
$w_{2,r'_2}$ & $0.472\pm0.009$ & Gaussian prior\\
$w_{2,K_p}$ & $0.375\pm0.011$ & Gaussian prior\\
\hline
\end{tabular}
\end{table}

Before creating MCMC chains, we first optimize free parameters for each light curve,
using the AMOEBA algorithm \citep{1992nrca.book.....P}.
The penalty function is given by
\begin{eqnarray*}
\label{eq:chisq}
\chi^2 &=& \sum_{\rm K2}\sum_{\rm t}\frac{(f_\mathrm{obs,t}-f_\mathrm{model,t})^2}{\sigma_{\rm f,t}^{2}}\\
&+& \sum_{\rm MuSCAT}\sum_{\rm t}\frac{(m_\mathrm{obs,t}-m_\mathrm{model,t})^2}{\sigma_{\rm m,t}^{2}}
+ \frac{(a/R_\star - 18.4)^2}{2.5^2},
\end{eqnarray*}
where $f_\mathrm{obs,t}$ and $\sigma_{\rm f,t}$ are
the relative fluxes of the target in each {\it K2} transit segment and their errors, and
$m_\mathrm{obs,t}$ and $\sigma_{\rm m,t}$ are
the magnitude of the target in $r'_s$ and $z_{x,2}$ bands and their errors.
The model functions $f_\mathrm{model,t}$ and $m_\mathrm{model,t}$
are combinations of the baseline model and the analytic transit formula mentioned above.
We note that the transit model is integrated over the {\it K2} cadence
($\sim 29.4$ minutes) for $f_\mathrm{model,t}$.
In addition, the time stamps of all data are converted to the $\mathrm{BJD_{TDB}}$ system
using the code by \citet{2010PASP..122..935E}.
If the reduced $\chi^2$ is larger than unity, we rescale the photometric errors of the data
such that the reduced $\chi^2$ for each light curve becomes unity.
We then estimate the level of time-correlated noise
(a.k.a. red noise: \cite{2006MNRAS.373..231P}) for each light curve,
by calculating the $\beta$ factor introduced by \citet{2008ApJ...683.1076W}.
The $\beta$ factor is used to take into account the time-correlated noise and
to properly compensate the possible underestimate of derived uncertainties
from analyses of transit photometry.
For the purpose, we compute the residuals for each light curve
and average the residuals into $M$ bins of $N$ points.
We then calculate the actual standard deviation of the binned data
$\sigma_{{\rm N,obs}}$ and the ideal standard deviation without any time-correlated noise
$\sigma_{{\rm N,ideal}} = \frac{\sigma_1}{\sqrt{N}} \sqrt{\frac{M}{M-1}}$,
where $\sigma_1$ is the standard deviation of the residuals for unbinned data.
To account for increased uncertainties due to the time-correlated noise,
we compute $\beta = \sigma_{{\rm N,obs}}/\sigma_{{\rm N,ideal}}$
for various $N$.
If $\beta$ is significantly higher than unity, it implies the presence of the time-correlated noise.
Consequently,
we find no significant time-correlated noise in {\it K2} light curves,
while MuSCAT light curves show significant time-correlated noise, especially in the $r'_2$ band.
We adopt $\beta = 1.56409$ for the $r'_2$ band and
$\beta = 1.08181$ for the $z_{s,2}$ band,
which are the median values of $\beta$ for $N=$ 5--20 binning cases in each band,
and further rescale the errors of the light curves by multiplying them by their $\beta$ factors.

Finally, we employ the Markov Chain Monte Carlo (MCMC) code
\citep{2013ApJ...773..144N,2016ApJ...819...27F}
to compute the posterior distributions for the free parameters.
We create 3 chains of 12,000,000 points,
and discard the first 2,000,000 points from each chain as ``burn-in''.
The jump sizes of parameters in each MCMC step are adjusted
such that acceptance ratios become $\sim$23\%,
which is considered as an optimal acceptance ratio for efficient convergence of MCMC
(see e.g., \cite{2005AJ....129.1706F}).

Table~\ref{planettable} presents the median values and uncertainties, which are defined
by the 15.87 and 84.13 percentile levels of the merged posterior distributions.
The baseline corrected transit light curves (in flux) are plotted in Figure~\ref{transitlc}.
We find that $R_p/R_\star$ for all three bands are consistent with one another
and that the MuSCAT light curves are consistent with a flat-bottomed transit.
To derive the orbital ephemeris we fit a linear model to the mid-transit times, 
yielding a period of $P=8.2669016 \pm 0.0000581$ days
and a time of first transit of $T_c (0) = 2457147.99107 \pm 0.00098$ (BJD$_{\rm TDB}$),
with $\chi^2$ of 11.499 for 8 degrees of freedom.
Note that in this fit we adopt the larger-side uncertainty if uncertainties of
respective mid-transit times are asymmetric.
To be conservative, we rescale the uncertainties of $P$ and $T_c (0)$ by $\sqrt{11.499/8}$,
and the rescaled uncertainties are presented in Table~\ref{planettable}.
This refined ephemeris will be useful for future transit observations of K2-105~b.

\begin{table}[tb]
\caption{Planetary Parameters of K2-105~b}\label{planettable}
\begin{tabular}{lc}
\hline
Parameter & Value \\\hline
{\it (MCMC Parameters)} & \\
$a/R_\star$ & $17.96_{-2.34}^{+0.91}$ \\
$b$ & $0.328_{-0.225}^{+0.249}$ \\
$R_p/R_\star$ [$K_p$ band]     & $0.03472_{-0.00067}^{+0.00133}$ \\
$R_p/R_\star$ [$r'_2$ band]      & $0.03444_{-0.00421}^{+0.00445}$ \\
$R_p/R_\star$ [$z_{s,2}$ band] & $0.03651_{-0.00541}^{+0.00388}$ \\
$T_{c} (0)$ [BJD - 2450000] & $ 7147.98960_{-0.00371}^{+0.00441} $\\
$T_{c} (1)$ [BJD - 2450000] & $ 7156.25371_{-0.00209}^{+0.00206} $\\
$T_{c} (2)$ [BJD - 2450000] & $ 7164.52415_{-0.00224}^{+0.00221} $\\
$T_{c} (3)$ [BJD - 2450000] & $ 7172.79050_{-0.00199}^{+0.00213} $\\
$T_{c} (4)$ [BJD - 2450000] & $ 7181.06464\pm0.00244 $\\
$T_{c} (5)$ [BJD - 2450000] & $ 7189.32677_{-0.00212}^{+0.00202} $\\
$T_{c} (6)$ [BJD - 2450000] & $ 7197.59274_{-0.00294}^{+0.00287} $\\
$T_{c} (7)$ [BJD - 2450000] & $ 7205.86050_{-0.00197}^{+0.00201} $\\
$T_{c} (8)$ [BJD - 2450000] & $ 7214.12742_{-0.00343}^{+0.00392} $\\
$T_{c} (34)$ [BJD - 2450000] & $ 7429.06529_{-0.00157}^{+0.00142}$\\
$K$ [m~s$^{-1}$] & $9.4 \pm 5.8$ ($<$26.8 \footnotemark[$*$])\\
\hline
{\it (Derived Parameters)} & \\
$P$ [days] & $8.266902 \pm 0.000070$ \\
$T_{c} (0)$ $^{\dagger}$ [BJD - 2450000] & $ 7147.99107\pm0.00117 $\\
$R_p$ [$R_{\oplus}$] \footnotemark[$\dagger$] & $ 3.59_{-0.39}^{+0.44} $ \\
$R_p$ [$R_{Jup}$] \footnotemark[$\ddagger$] & $0.369_{-0.034}^{+0.039} $ \\
$i$ [$^\circ$] & $88.95_{-1.07}^{+0.73}$ \\
$T_{14}$ [days] &  $0.14426_{-0.00203}^{+0.00224}$ \\ 
$M_p$ [$M_{\oplus}$] & $30 \pm 19$ ($<$90 \footnotemark[$*$])\\
\hline
\end{tabular}
\begin{tabnote}
\hangindent6pt\noindent
\hbox to6pt{\footnotemark[$*$]\hss}\unskip An upper limit at 99.865 percentile (3$\sigma$) level.
\hbox to6pt{\footnotemark[$\dagger$]\hss}\unskip This is the origin for the transit ephemeris.
\hbox to6pt{\footnotemark[$\ddagger$]\hss}\unskip Based on $R_p/R_\star$ in $K_p$ band.
\end{tabnote}
\end{table}

\subsection{Subaru/HDS Radial Velocities and a Mass Upper Limit}

	\begin{figure}[tp]
			\includegraphics[width=8.5cm]{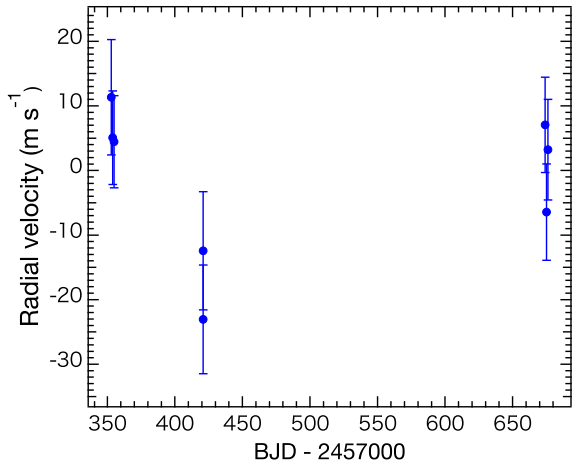} 
				\caption{Relative radial velocities (blue points) of K2-105 taken with the Subaru HDS.
				}
			\label{fig_rvjd}
	\end{figure}

	\begin{figure}[tp]
			\includegraphics[width=8.5cm]{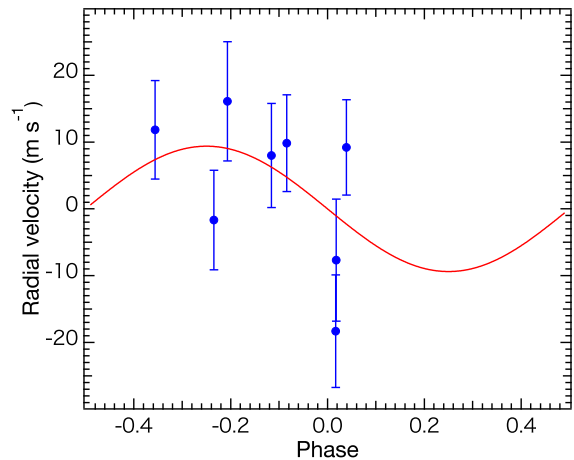} 
				\caption{Radial velocities (blue points) phased by the orbital period given in Table~\ref{planettable}
				and the best-fit RV model (red line).
				}
			\label{fig_rvphase}
	\end{figure}

We employ the RV pipeline for the Subaru HDS described in \citet{2002PASJ...54..873S}
to extract the relative RVs with respect to the template iodine-free spectrum.
The derived RVs are presented in Table~\ref{rvtable} and plotted in Figure~\ref{fig_rvjd}.
We do not find any significant long-term radial velocity drift.

To model the observed RVs, we adopt an RV model,
$v_\mathrm{model}=-K \sin \phi+\gamma$, where $K$, $\phi$, $\gamma$ are 
the RV semi-amplitude, the orbital phase relative to the mid-transit,
and the offset RV relative to the template spectrum.
We fix the orbital period $P$ to 8.2669016 days and the origin of the transit ephemeris
$T_c (0)$ to 2457147.99107 (BJD$_{\rm TDB}$) as derived in \S3.3.
We do not consider the Rossiter-McLaughlin effect, since there is no RV data during a transit.
We also neglect the eccentricity $e$, since it is indeterminable with the current RVs.

We first optimize $K$ and $\gamma$ using the AMOEBA algorithm \citep{1992nrca.book.....P},
and create an MCMC chain of 500,000 points starting from the optimal parameters.
The acceptance ratio is set to $\sim$23\%.
The phased RVs and the best-fit RV model are shown in Figure~\ref{fig_rvphase}.
The median values and uncertainties of the free parameters
are presented in Table~\ref{planettable}.
We also present a 3$\sigma$ upper limit (99.865 percentile level) of $K$
in Table~\ref{planettable}.
Consequently, the RV semi-amplitude is $9.4\pm5.8$ m~s$^{-1}$,
which corresponds to $30\pm19$ M$_{\oplus}$ for the mass of K2-105~b.
At this point, the current RVs are not sufficient to determine 
the RV semi-amplitude and the mass of K2-105~b precisely.
Nevertheless, we can put a constraint on $K < 26.8$ m~s$^{-1}$ at the 3$\sigma$ level,
which corresponds to a mass upper limit of $90 M_{\oplus}$ or  $0.00027 M_{\odot}$.
This upper limit ensures that the mass of K2-105~b is within a planetary mass,
excluding the possibility of an eclipsing binary scenario for this system.

\begin{table}[t]
\caption{Radial velocities of K2-105 taken with Subaru/HDS.\label{rvtable}}
\begin{tabular}{lrr}
\hline
BJD$_{\rm TDB}$ & Value [m s$^{-1}$] & Error [m s$^{-1}$]  \\\hline
2457352.95525 & 11.33 & 8.92 \\
2457353.96869 & 5.08   & 7.25 \\
2457354.98800 & 4.45   & 7.14 \\
2457420.93976 & -23.06 & 8.42 \\
2457420.94741 & -12.43 & 9.14 \\
2457674.12841 & 7.07   & 7.38 \\
2457675.13374 & -6.44  & 7.45 \\
2457676.11418 & 3.23   & 7.80 \\
\hline
\end{tabular}
\end{table}

\section{Discussions}

\subsection{Confirmation of the Planetary Nature of K2-105~b}

To further validate the planetary nature of the transit signal,
we use the open-source Python code \texttt{vespa} 
\citep{2012ApJ...761....6M, 2015ascl.soft03011M}, which employs a 
robust statistical framework to calculate
the False Positive Probability (FPP) of the transit signal.
It does this by taking into account a variety of factors: 
the size of the photometric aperture of {\it K2}, the
source density along the line of sight as determined from galaxy simulations,
constraints on contaminants from high resolution imaging contrast curves,
physical properties of the host star from spectroscopically-derived parameters
and broadband photometry, and comparisons of the shape of the phase-folded {\it K2} light curve
to a large number of realistic false positive scenarios. 

We input our results presented in the last section to \texttt{vespa} and find 
the final FPP for this target to be extremely low ($<10^{-6}$),
which strongly indicates a planetary nature for the origin of the observed transit signals.
We therefore rule out all of the false positive scenarios accounted for by \texttt{vespa}
(i.e. hierarchical triple systems, eclipsing binaries, blended background eclipsing binaries).

We conclude that K2-105~b is a bona fide planet,
based on the mass constraint presented in \S3.4 and the extremely low FPP.

\subsection{The New Transit Ephemeris and a Hint of Transit Timing Variation}

We have derived the new transit ephemeris for K2-105~b as
$P = 8.266902 \pm 0.000070$ days and 
$T_{c} (0) = 2457147.99107\pm0.00117$ in BJD$_{\rm TDB}$.
This is indeed the first reliable transit ephemeris for K2-105~b,
since \citet{2016MNRAS.461.3399P} and \citet{2016arXiv160702339B},
who reported EPIC~211525389~b as a candidate planet, presented
a transit ephemeris without any uncertainty.
Using the transit ephemeris we have derived from our observations,
transits of K2-105~b in 2017 can be predicted with
uncertainties of only about 10 minutes, which will facilitate
the scheduling of future transit observations.

\begin{figure}[tp]
			\includegraphics[width=8.5cm]{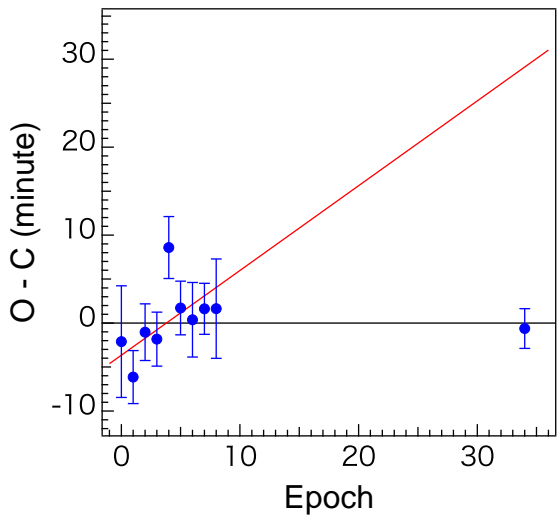} 
				\caption{Residuals of the observed mid-transit times from the current best-fit transit ephemeris.
				The red line represents the best-fit transit ephemeris for the transits from {\it K2} only.
				We note that the transit ephemeris from {\it K2} has an uncertainty of $\sim$17 minutes
				at the epoch of 34. Thus the discrepancy is still within the 2$\sigma$ level.
				}
			\label{fig_ttv}
	\end{figure}

We check the possible presence of transit timing variation (TTV)
for the transits of K2-105~b.
Figure~\ref{fig_ttv} plots residuals of the observed mid-transit times
from the current transit ephemeris.
While if only {\it K2} transits are taken into account, a linear fit to the mid-transit times
gives $P = 8.2675710 \pm 0.0003569$ days and
$T_c (0) = 2457147.98853 \pm 0.00164$ in BJD$_{\rm TDB}$,
with $\chi^2$ of 7.884 for 7 degree-of-freedom.
The transit for the MuSCAT run
occurred about 30 min earlier than the prediction from the {\it K2}-only transits,
although the difference is at the 2$\sigma$ level.
The discrepancy is statistically not significant, but
it may suggest that an additional non-transiting planet exist as is
the case for K2-19 b \& c (e.g., \cite{2015A&A...582A..33A,2015ApJ...815...47N}).
Alternatively, stellar activities such as star spots may play a role in the apparent discrepancy
(see e.g., \cite{2013A&A...556A..19O}).
To confirm the presence of TTVs for K2-105~b,
further transit monitoring is needed.

\subsection{K2-105~b in the Context of Period-Radius and Mass-Radius Relation}

\begin{figure}[tp]
			\includegraphics[width=9cm]{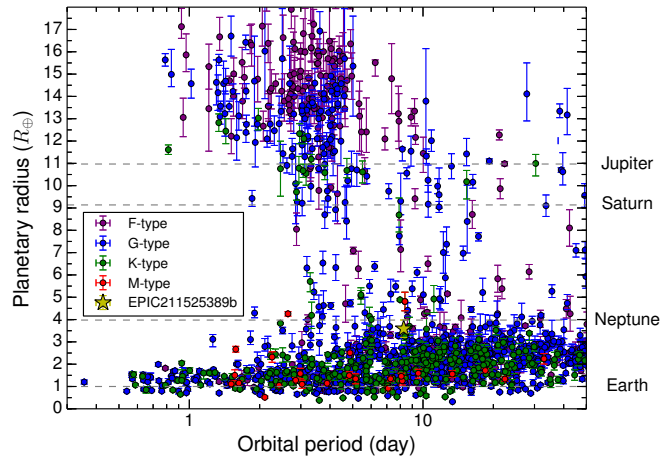} 
				\caption{Period-radius relation of confirmed transiting exoplanets with
				$P \leq 50$\,days around F-type ($M_\star = 1.04 - 1.4\,M_\odot$),
				G-type ($M_\star = 0.8 -1.04\,M_\odot$), K-type ($M_\star = 0.45 -0.8\,M_\odot$),
				and M-type stars ($M_\star = 0.08 -0.45\,M_\odot$) as of 2016 August;
				the data come from http://exoplanet.eu.
				Planets for which the radius uncertainty exceeds 20\% of their representative values are excluded.}
			\label{fig_PR}
	\end{figure}

Occurrence rates of planets derived from RV surveys and
the {\it Kepler} indicate that short-period Neptune-sized planets such as K2-105~b
are only rarely found in planetary systems around solar-type stars
(e.g., \cite{2012ApJS..201...15H,2013PNAS..11019273P}).
Figure~\ref{fig_PR} shows an orbital period--radius distribution of transiting planets with $P \leq 50$\,days. 
We see a clear lack of planets with radii of $\sim 3.5 - 10\,R_\oplus$
around solar-type stars, albeit interestingly, no hot Jupiter with radius of
$\gtrsim 10\,R_\oplus$ is seen around stars with $M_\star \leq 0.45\,M_\odot$.
These features, also pointed out by previous studies
(e.g., \cite{2016A&A...589A..75M,2016ApJ...820L...8M}),
may reflect a size boundary between a failed core and a gas giant,
which corresponds to a critical core mass that triggers gas accretion in a runaway fashion,
or mass loss via photo-evaporation.
The latter case can be a useful indicator to evaluate the efficiency of atmospheric escape
due to a stellar irradiation or injection of high-energy particles.
Thus, the discovery of K2-105~b can be an interesting benchmark
to disentangle the origin of Neptune-sized planets close to central stars.

\begin{figure}[tp]
			\includegraphics[width=8.1cm]{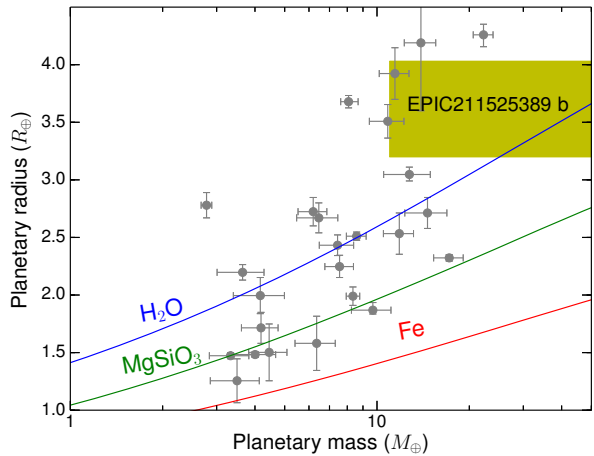} 
				\caption{Mass-radius relation of confirmed transiting exoplanets as of 2016 September; the data come from http://exoplanet.eu. Planets with uncertainties in mass and radius over 20\% of their representative values are not shown here. Theoretical models of iron, water, and silicate planets are based on \citet{2013PASP..125..227Z}. We adopt the possible range of K2-105~b's mass derived from RV measurements by the Subaru HDS, $M_{\rm p} = 30\pm19\,M_\oplus$ and its radius of $R_{\rm p} = 3.59^{+0.44}_{-0.39}\,R_\oplus$.}
			\label{fig_MR}
	\end{figure}

Figure~\ref{fig_MR} shows theoretical mass-radius relations for three types of planets
and transiting exoplanets with known mass.
We find that K2-105~b is not a bare rocky planet but likely has
an atmosphere ($<$ 10\% of its total mass) if its total mass is smaller than $30\,M_{\oplus}$.
K2-105~b orbits at $a_{\rm p} = 0.081 \pm 0.006$\,AU around a G-dwarf with
the mass of $1.01 \pm 0.07\,M_\odot$.
According to \citet{2013ApJ...775..105O}, K2-105~b can retain its atmosphere
under an intense stellar X-ray and EUV irradiation during an estimated stellar age of
older than 0.6 Gyr, if the core mass is greater than $\sim 6\,M_\oplus$.

If K2-105~b is a gas dwarf, how did it form?
There are two possible formation scenarios, namely, in-situ gas accretion onto a massive core
\citep{2012ApJ...753...66I,2014ApJ...797...95L,2015MNRAS.447.3512O} or inward migration of
a Neptune-like planet (e.g., \cite{2014ApJ...791..103B}).
However, we cannot rule out both stories because of an unknown mass of K2-105~b.
Thus, mass determination from follow-up RV observations will be indispensable for 
constraining the formation history and quantifying the effect of photo-evaporation.

In addition, close-in Neptune-sized planets as represented by K2-105~b would be
suggestive of uncovering how the Solar System was born.
There is no K2-105~b-like planet in the Solar System,
instead the two ice giants orbit beyond $\sim$20\,AU.
As one possibility, this might be caused by the presence of Jupiter and Saturn orbiting
within the orbits of the two ice giants, acting as a barrier against inward migrating cores.
Long-term RV monitoring of K2-105 to constrain the possibility of outer giant planets
should be helpful in understanding the orbital evolutions of Neptune-like planets and the Solar System.
Therefore, long-term RV monitoring of this system would be also encouraged.

\section{Summary}

We have confirmed the planetary nature of K2-105~b,
using transit photometry from the {\it K2} mission,
high dispersion spectroscopy and RVs from Subaru/HDS,
high-contrast AO imaging from Subaru/HiCIAO,
and ground-based transit photometry from Okayama/MuSCAT.
The host star K2-105 is located in {\it K2} campaign field 5,
and estimated to be a metal-rich G-dwarf.
Although further RV monitoring is required to precisely determine the mass of K2-105~b,
the Subaru HDS RVs put a stringent constraint on the mass of K2-105~b
as less than 90$M_\oplus$ or  $0.00027M_{\odot}$ at the 3$\sigma$ level,
ensuring that the mass of K2-105~b is well within the planetary mass range.
Our joint analysis of the transit data from {\it K2} and MuSCAT
yields an orbital period of $P  = 8.266902 \pm 0.000070$ days and
an origin of mid-transit time $T_{c} (0) = 2457147.99107\pm0.00117$ in BJD$_{\rm TDB}$.
The transit ephemeris is accurate enough to predict transit times of K2-105~b
with uncertainties of less than 20 minutes for the next few years.
We have found that the transit observed with the Okayama/MuSCAT occurred
about 30 minutes earlier than the prediction from the {\it K2}-only transits.
Although the discrepancy from the prediction is statistically marginal
at the 2$\sigma$ level,
this may suggest that additional long period or non-transiting planet(s) exist in the system,
which increases the need for further transit and RV measurements of this system.

The transit depth of K2-105~b, $R_p/R_\star \sim 0.035$, corresponds to
a planetary radius of $R_p  = 3.59_{-0.39}^{+0.44}$ $R_{\oplus}$.
Thus K2-105~b is a short-period Neptune-sized planet.
As K2-105~b is a transiting planet around a relatively bright host star,
it is a favorable and important target for characterization
of its mass via RV measurements, its atmosphere via transmission spectroscopy, 
spin-orbit (mis)alignment via the Rossiter-McLaughlin effect or doppler tomography,
and the presence of additional planet via TTVs and/or RV trends.
Such further characterization will be vital for understanding
the formation and migration history of this planetary system.

\section{Funding}

This work was supported by Japan Society for Promotion of Science (JSPS)
KAKENHI Grant Numbers JP25247026, JP16K17660, 25-8826, JP16K17671, and JP15H02063.
This work was also supported by the Astrobiology Center Project of
National Institutes of Natural Sciences (NINS)
(Grant Numbers AB271009, AB281012 and JY280092).
I.R. acknowledges support by the Spanish Ministry of Economy
and Competitiveness (MINECO) through grant ESP2014-57495-C2-2-R.

\begin{ack}
We acknowledge Roberto Sanchis-Ojeda, who established the ESPRINT collaboration.
We thank supports by Akito Tajitsu and Hikaru Nagumo for our Subaru HDS observation,
Jun Hashimoto for Subaru HiCIAO observation, and Timothy Brandt for HiCIAO data reduction.
This paper is based on data collected at the Subaru telescope and Okayama 188cm telescope,
which are operated by the National Astronomical Observatory of Japan.  
The data analysis was in part carried out on common use data analysis computer system
at the Astronomy Data Center, ADC, of the National Astronomical Observatory of Japan.
PyFITS and PyRAF were useful for our data reductions.
PyFITS and PyRAF are products of the Space Telescope Science Institute, which is operated by AURA for NASA.  
Our analysis is also based on observations made with the NASA/ESA Hubble Space Telescope,
and obtained from the Hubble Legacy Archive, which is a collaboration between
the Space Telescope Science Institute, the Space Telescope European Coordinating Facility (ST-ECF/ESA)
and the Canadian Astronomy Data Centre (CADC/NRC/CSA).
This work has made use of data from the European Space Agency (ESA)
mission {\it Gaia} (http://www.cosmos.esa.int/gaia), processed by
the {\it Gaia} Data Processing and Analysis Consortium (DPAC,
http://www.cosmos.esa.int/web/gaia/dpac/consortium). Funding
for the DPAC has been provided by national institutions, in particular
the institutions participating in the {\it Gaia} Multilateral Agreement.
We acknowledge the very significant cultural role and reverence that the
summit of Mauna Kea has always had within the indigenous people in Hawai'i. 
\end{ack}



\end{document}